\documentclass[preprint,preprintnumbers,amsmath,amssymb,floatfix]{revtex4}

\usepackage{graphicx}
\usepackage{dcolumn}
\usepackage{bm}
\usepackage{epsfig}

\begin{document}

\centerline{\normalsize DESY 14--187 \hfill ISSN 0418--9833}

\title{Production of massless charm jets in pp collisions at next-to-leading
order of QCD}

\author{Isabella Bierenbaum}

\email{isabella.bierenbaum@desy.de}
\altaffiliation[Current address: ]{Institutes of Physics and Mathematics, Humboldt-Universit\"at zu Berlin, Unter den Linden 6, 10099 Berlin, Germany}

\author{Gustav Kramer}

\email{gustav.kramer@desy.de}

\affiliation{{II.} Institut f\"ur Theoretische Physik,
Universit\"at Hamburg, Luruper Chaussee 149, 22761 Hamburg, Germany}

\date{\today}

\begin{abstract}

We present predictions for the inclusive production of charm jets in
proton-proton collisions at 7 TeV. Several CTEQ parton distribution
functions (PDFs) of the CTEQ6.6M type are employed, where two of the
CTEQ6.6 PDFs have intrinsic charm. At large enough jet transverse
momentum and large jet rapidity, the intrinsic charm content can be
tested.

\end{abstract}

\maketitle
\thispagestyle{empty}

\section{Introduction}

The cross section for producing heavy quarks in proton-proton
collisions at the Large Hadron Collider (LHC) can be computed in
QCD. Calculations have been carried out at next-to-leading order (NLO)
in perturbation theory \cite{1}.  So far, most of the predictions for
charm-quark jets have been done in the massive quark scheme or
fixed-flavor-number scheme (FFNS) in which charm quarks appear in the
final state only and not as partons in the initial state, as for
example in the zero-mass variable-flavor-number scheme (ZM-VFNS). The
treatment of the charm quark being massless is justified as long as
the transverse momentum $p_T$ of the produced jets is large enough,
i.e. for $p_T^2 \gg m^2$, where $m$ is the charm-quark mass. In such a
case, one might hope to obtain information on the charm content of the
proton parton distribution function (PDF) by measuring cross sections
in pp collisions for the production of charm jets at large $p_T$. Such
a task has been pursued in connection with dijet photoproduction of
charm jets by M. Klasen and one of us in order to get information on
the photon PDF \cite{2}.  Another way to investigate the charm PDFs is
to measure, instead of charm jets, the production of charmed hadrons,
as for example the various kinds of D mesons or charmed baryons. This
has been considered in several publications on the production of D
mesons in photoproduction and hadron-hadron collisions in the ZM-VFNS
\cite{3,4} and the general-mass variable-flavor-number scheme
(GM-VFNS) \cite{5}. The GM-VFNS incorporates finite charm-mass
corrections taken from the FFNS in order to improve the predictions at
small and medium $p_T$s.  This approach has already been used to
predict single D-meson inclusive cross sections for various LHC
experiments \cite{6}, which are found in good agreement with the
experimental data \cite{7,8,9}. Of course, these cross sections depend
on the fragmentation functions for the fragmentation of the final
state partons into the respective charmed mesons D or charmed baryons
which are the result of fits to production cross sections at $e^+e^-$
colliders. On the other hand, the charm-jet production cross sections
do not depend on such fragmentation functions. Thus we can expect to
obtain complementary information on the charm PDFs to the
charmed-hadron cross sections, independent of the fragmentation
functions input. The only requirement, however, is that the selection
of charmed jets in the measurement must correspond to the selection in
the theoretical cross section calculations.  In the next section, we
shall describe the theoretical framework and outline the input in
terms of PDFs for the initial state. Section 3 contains our results
for the charm-jet cross section. In this section, we also show a
comparison with the single-inclusive jet cross section measured by the
CMS collaboration at the LHC \cite{10} in order to demonstrate that
our jet cross section routine agrees with the experimental data with
no special flavor selection. Similar data are also obtained by the
ATLAS collaboration at the LHC \cite{10a}. A summary and our
conclusions are presented in Section 4.

\section{Theoretical framework and PDF input}

For our calculation, we rely on previous work on dijet production in
the reaction $\gamma + p \rightarrow jet + X$ \cite{11}, in which
cross sections for inclusive one-jet and two-jet production up to NLO
for both the direct and the resolved contributions are calculated (for
a review see \cite{12}). The predictions of this work have been tested
by many experimental studies of the H1 and ZEUS collaboration at
HERA. The resolved part of this cross section routine can be used for
pp collisions replacing the photon PDF by the proton PDF. Recent
comparisons of the Tevatron and LHC jet data are usually performed
with predictions of the NLOJET++ routine \cite{13} within the
framework of FASTNLO \cite{14}. The routine \cite{11} contains quarks
of all flavors and the gluon.  For our purposes, it has been restricted
to the case that at least one charm quark appears in the final
state. In the initial state we have the contributions $cg$,
$\bar{c}g$, $cq$, $\bar{c}q$, $c\bar{c}$, $cc$, $\bar{c}\bar{c}$,
where $q$ is a light quark (or antiquark), with the restriction that
$c\bar{c} \rightarrow gg$ and $c\bar{c} \rightarrow q\bar{q}$ in
leading order (LO) and the corresponding contributions in NLO are
removed, i.e. only terms like $c\bar{c} \rightarrow c\bar{c}$ and $cc
\rightarrow cc$, $\bar{c}\bar{c} \rightarrow \bar{c}\bar{c}$
respectively, are retained.  In addition, there are also contributions
with light quarks and gluons in the initial state, as for example
$q\bar{q} \rightarrow c\bar{c}$ and $gg \rightarrow c\bar{c}$, as well
as the corresponding NLO contributions.

For our predictions, we employ various PDFs of the proton. For the
inclusive jet cross section containing all flavors and to be compared
with CMS measurements \cite{10}, we use the CTEQ CT10 version
\cite{15} as used in the CMS publication \cite{10} with $n_f=5$,
i.e. we include all flavors up to including the bottom quark. The
asymptotic scale parameter is $\Lambda_{\overline{MS}}^{(5)} = 0.262$
GeV corresponding to $\alpha_s^{5}(m_Z) =0.118$. We choose the
renormalization scale $\mu_R = \xi_R p_T$ and the factorization scale
$\mu_F=\xi_F p_T$, where $p_T$ is the largest transverse momentum of
the two (or three) final state jets. $\xi_R$ and $\xi_F$ are
dimensionless scale factors, which are varied about their default
values $\xi_R = \xi_F=1$ to be specified later. The center-of-mass
(c.m.) energies of the proton in all calculations are taken as
$\sqrt{S} = 7$ TeV, as for the data of CMS \cite{10}. The bin size in
$p_T$ is taken according to the CMS publication \cite{10} and the
rapidity interval is $-0.5 \leq y \leq 0.5$ as given in one of the
curves in \cite{10}. For the charm-jet cross sections, we concentrate
our calculations on the large $p_T$ region, $p_T \geq 37$ GeV. The bin
sizes in $p_T$ and two of the rapidity bins ($|y| \leq 0.5$ and $2.0
\leq |y| \leq 2.2$) for our charm-jet calculation are chosen in
accordance with the CMS publication for inclusive b-jet production
\cite{16} since there are no data for inclusive c-jet production
yet. In addition, we select as a third rapidity interval: $2.2 \leq
|y| \leq 3.2$. The charm-jet calculations are done with the CTEQ6.6M
PDF sets \cite{17} and $n_f$ is taken to be $n_f=4$. The corresponding
$\alpha_s$ is calculated with $n_f=4$ as well, with the $\Lambda$
value corresponding to this $n_f$ value. In the theoretical
calculations, jets were reconstructed with the $k_T$-cluster algorithm
in the longitudinally invariant inclusive mode \cite{18}. The
measurements of the inclusive jet cross sections are done with the
anti-$k_T$ jet clustering algorithm \cite{19} using as the distance
parameter R=0.5. This value for R is also applied for all the other
calculations in this work. At $O(\alpha_s^2)$, the parton-level
predictions from the $k_T$ and anti-$k_T$ algorithms are identical.

\section{Results}

First we show our prediction of $d\sigma/dp_T$ for the $p+p
\rightarrow single~jet+X$ cross section in the $p_T$ range $18 \leq
p_T \leq 1684$ GeV and $|y| \leq 0.5$ in $p_T$ bins as chosen by CMS
\cite{10} for their measurement. Non-perturbative (NP) corrections for
hadronization and multiple parton interactions were estimated by CMS
\cite{10} and are published in the Durham HepData project
\cite{20}. They are applied to our NLO perturbative QCD
predictions. For low-$p_T$ jets, the NP corrections are as large as
$40\%$ with a relative uncertainty of $91\%$ and for large-$p_T$ jets
they are around $1\%$ with an error of $0.1\%$.  Theoretical errors
from the dependence on the choice of the renormalization scale $\mu_R$
and the factorization scale $\mu_F$ are determined by varying the
scales according to the following combinations of $\xi_R$ and $\xi_F$:
(1/2,1/2), (1/2,1), (1,1/2), (1,2), (2,1) and (2,2), and the largest up
and down cross sections for these choices are taken as the scale
variation. The default choice
\begin{figure*}
\includegraphics[width=7.5cm]{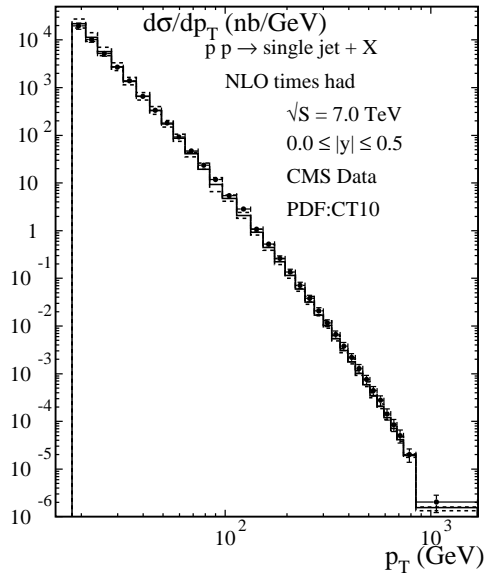}
\caption{\label{fig:1} Single-inclusive jet cross sections
  $d\sigma/dp_T$ as a function of $p_T$ compared to the data from CMS
  \cite{10}. The NLO theoretical predictions are corrected for
  non-perturbative effects via multiplicative factors. The theoretical
  error is obtained by independent scale variations given as the
  dashed curves. The solid curve indicates the default scale choice.}
\end{figure*}
is (1,1) with $\mu_R=\mu_F=p_T$. These scale variations modify the
prediction of the inclusive jet cross section at the lowest $p_T$ bin
($18-21$ GeV) by $+26.0\% (-18.7\%)$, the medium $p_T$ bins ($37-43$
GeV) by $+19.2\%(-16.5\%)$, ($97-114$ GeV) by $+15.1\%(-12.8\%)$, and
the two large $p_T$ bins ($174-196$ GeV) by $+13.1\%(-11.8\%)$ and bin
($846-1684$ GeV) by $+4.4\%(-13.0\%)$. The second and fourth value
will be used later for estimates of charm-jet cross section
errors. The corrected inclusive jet cross section including the
theoretical error together with the CMS data \cite{10} is presented in
Fig.~1. It shows the jet $p_T$ spectra between $18$ and $1684$ GeV.
The agreement between data and the theoretical prediction is quite
good and quite similar to the comparison shown in the CMS paper
\cite{10}. This reference also contains cross section data and a
comparison with theory predictions for five more $|y|$ bins up to
$|y|=3$, which we have not calculated, exhibiting a similar good
agreement between data and pQCD calculations.

\begin{figure*}
\includegraphics[width=7.5cm]{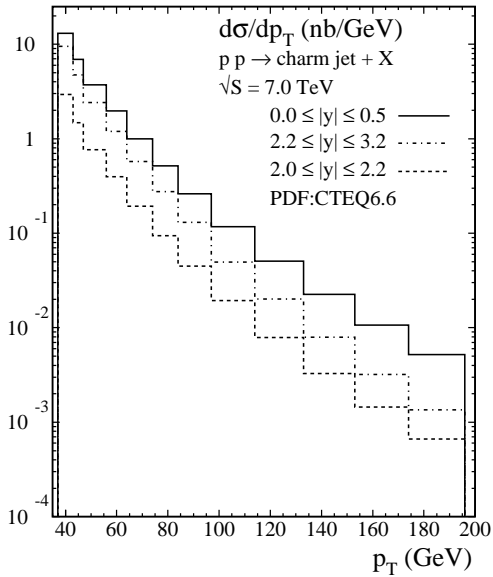}
\caption{\label{fig:2} Cross section $d\sigma/dp_t$ for three $|y|$
  regions as a function of $p_T$ for various bins at larger $p_T$.}
\end{figure*}
The cross section for inclusive charm-jet production is shown in
Fig.~2. These cross sections are calculated in 12 $p_T$ bins for $p_T$
between $p_T=37$ GeV and $p_T=196$ GeV, using the CTEQ6.6M PDF set
\cite{17}. The widths of the bins have been chosen according to the
CMS publication for b-jet production \cite{16}, since so far c-jet
production cross sections, which could guide us, have not been
measured. The cross sections $d\sigma/dp_T$ have been calculated for
three rapidity bins: $0.0 \leq |y| \leq 0.5$, $2.0 \leq |y| \leq 2.2$
and $2.2 \leq |y| \leq 3.2$. The first and second bin are taken as in
\cite{16}, while the third $|y|$ bin was chosen for the purpose of
comparing with cross sections with intrinsic charm PDFs to be
considered later. In Fig.~2, the largest cross section as a function of
$p_T$ appears for $|y| \leq 0.5$. The second largest is the one for
$2.2 \leq |y| \leq 3.2$, which is larger than the cross section for
$2.0 \leq |y| \leq 2.2$ due to the bigger $|y|$-bin size. All cross
sections in the three $|y|$ bins have a similar dependence on $p_T$.
The contribution of the c-jet to the full inclusive jet cross section
lies between $1.5\%$ and $2\%$, which is shown in Fig.~3, where we
have plotted the ratio of the cross sections c-jet/inclusive jet for
the three $|y|$ bins. For the $|y|$ bin: $0.0 \leq |y| \leq 0.5$ this
ratio is slightly larger than for the other two $|y|$ bins. (For
better visibility the ratios for $2.0 \leq |y| \leq 2.2$ and $0.0 \leq
|y| \leq 0.5$ are multiplied by factors 10 and 100, respectively). The
curves in Fig.~2 and 3 are our main result. We see that the charm
cross section in relation to the inclusive jet cross section is rather
small. The plots in Fig.~2 and Fig.~3 do not contain non-perturbative
corrections for hadronization and multiple parton interactions. If
they would be the same as in the inclusive jet cross sections, this
would not change the ratio in Fig.~3. In the considered $p_T$ range
and $|y| \leq 0.5$ these corrections are between $12\%$ and $3\%$ (for
increasing $p_T$ for the bins between $p_T=37$ GeV and $p_T=196$ GeV)
in the case of the inclusive jet cross sections \cite{20}. We expect
them to be somewhat larger for c-jets than for inclusive
jets. However, they have not been calculated yet. Such NP corrections
could be estimated using PYTHIA or HERWIG Monte Carlo programs as has
been done in the inclusive jet case \cite{10,20}.
Another issue is the selection of the jet events on the experimental
and the theoretical side. On the experimental side, we assume that the
c-jets are identified by finding the secondary decay vertex of the
c-hadrons. This means that at least one charmed hadron is in the final
state. The inclusive charm-jet cross section $d\sigma/dp_T$ contains
in addition also a gluon jet or light quark jet, although of course
not a light-quark or b-jet instead of the c-jet. For example, already
in LO the process $cg \rightarrow cg$ leads to a c-jet and a gluon jet
which both contribute to $d\sigma/dp_T$ and similarly at NLO.
\begin{figure*}
\includegraphics[width=7.5cm]{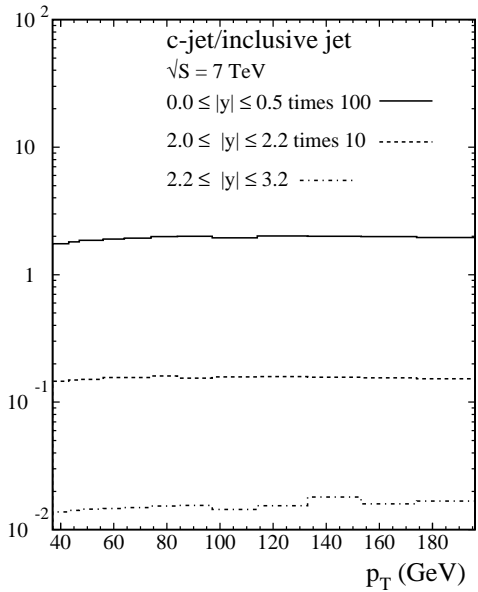}
\caption{\label{fig:3} Ratio of single-inclusive charm-jet cross
  section and the single-inclusive jet cross section as a function of
  $p_T$ for three rapidity regions $|y|$. The ratios for the lowest
  two $|y|$ regions are multiplied by 10 and 100.}
\end{figure*}
In all common proton PDFs the charm content is generated
perturbatively, i.e.  by assuming that at threshold near $\mu_F = m$
the PDF vanishes or is small and given by higher order contributions
while at larger scales $\mu_F$ it is determined by evolution. Since
many decades it is argued that the charm PDF $c(x,\mu_F)$ at $x > 0.1$
could have a non-perturbative intrinsic charm contribution due to the
fact that the charm mass is not really large compared to the QCD scale
parameter $\Lambda$. In recent work by the CTEQ collaboration, it was
investigated how much of the intrinsic charm inside specified models
is compatible with the global data samples for particular PDF
parametrizations.  Such studies were performed starting from the
parametrization CTEQ6.5 in \cite{22}, starting from CTEQ6.6 in
\cite{17} and quite recently starting from CTEQC10 in \cite{23}. In
all three investigations \cite{22,17,23}, two intrinsic charm models
have been used: (i) a valence-like parton distribution (BHPS model)
and (ii) a sea-like parton distribution (SEA model). We select these
two models BHPS and SEA, (see \cite{22} for details) with the
constraint obtained in the CTEQ analysis \cite{17}. We prefer the
corresponding CTEQ6.6c PDF in the program library LHAPDF \cite{24}
with a $3.5\%$ $(c+\bar{c})$
\begin{figure*}
\includegraphics[width=7.5cm]{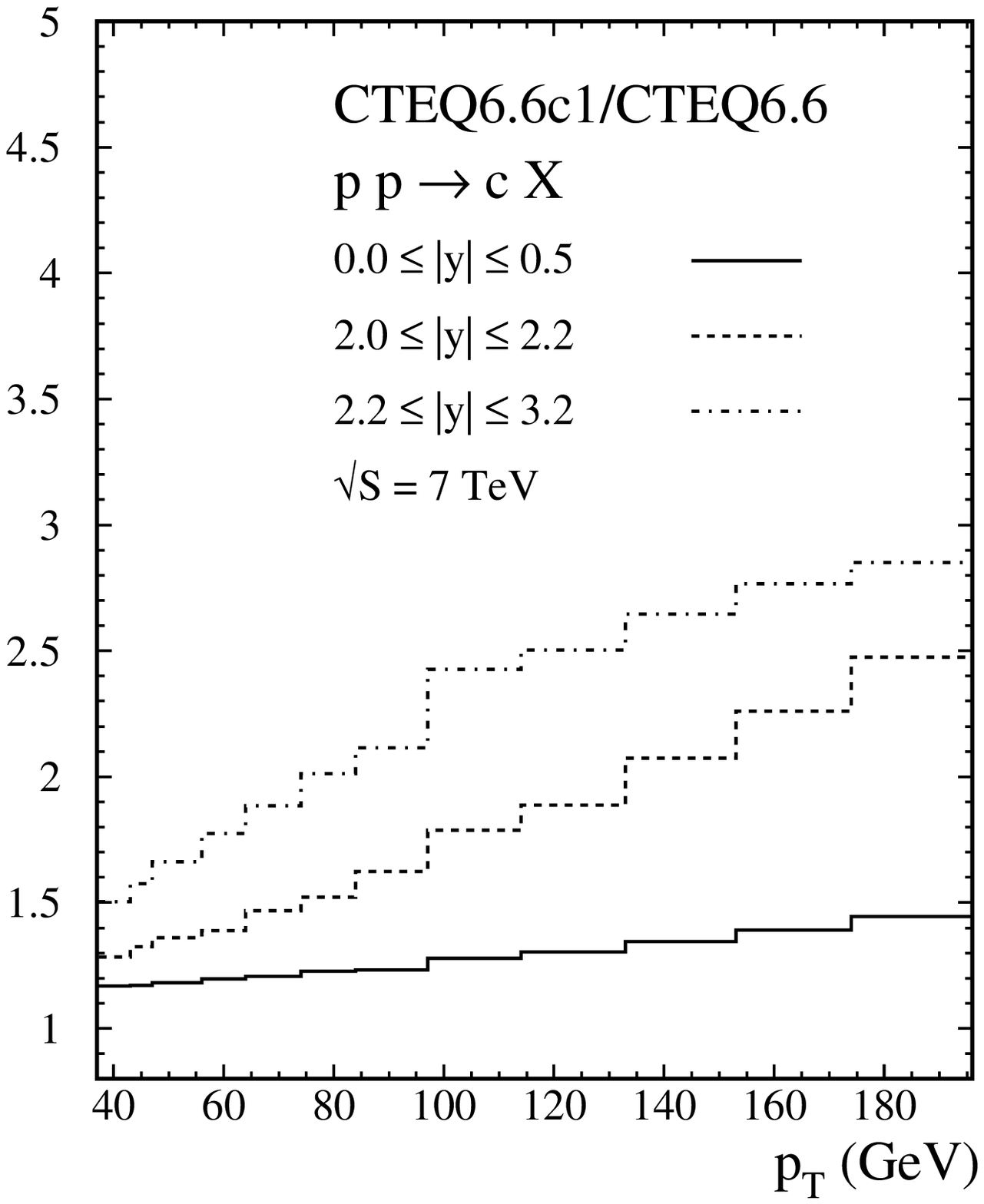}
\includegraphics[width=7.5cm]{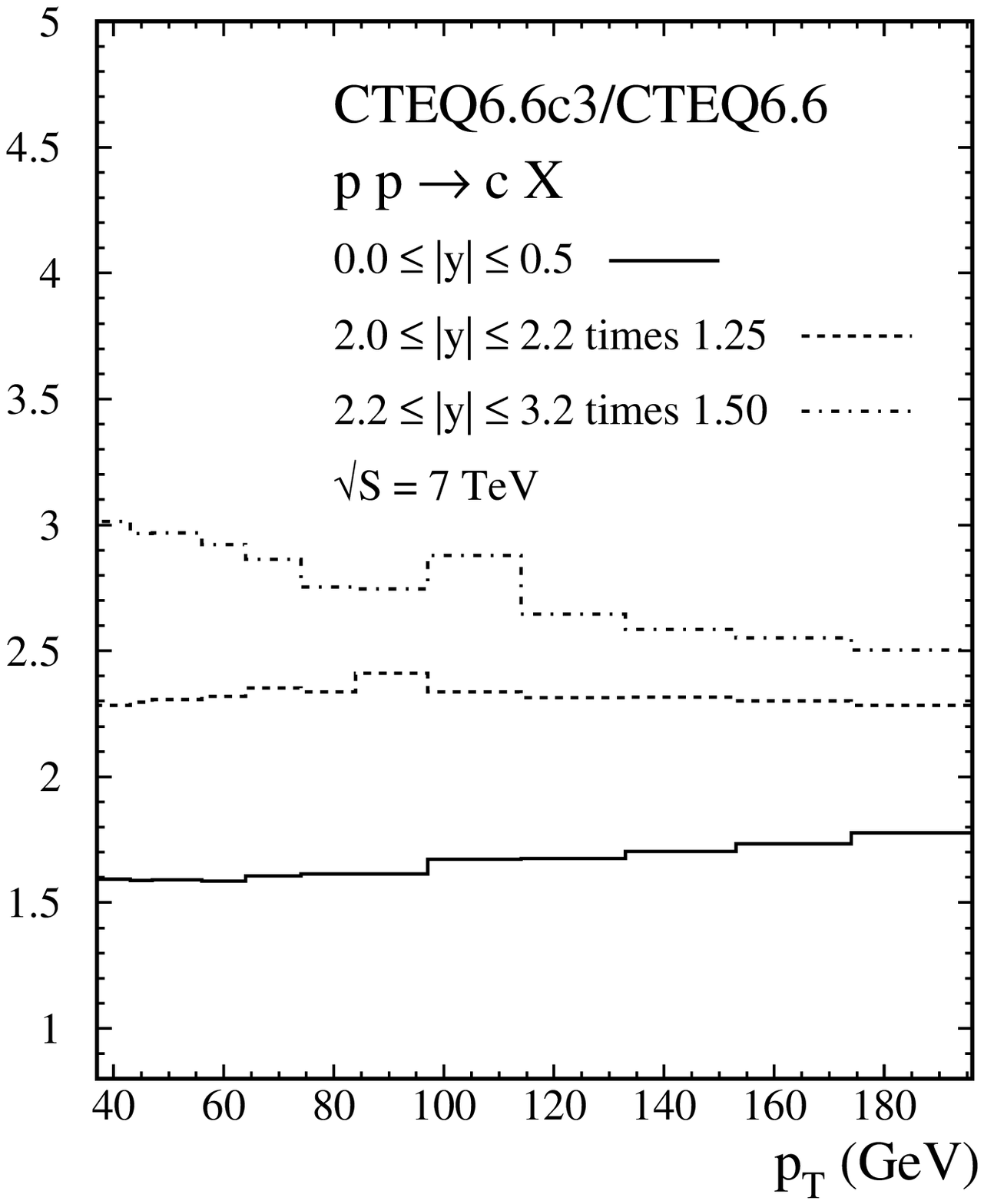}
\caption{\label{fig:4} Ratio of single-inclusive charm-jet cross
  section for intrinsic charm with BHPS (left-hand side) and SEA
  (right-hand side) modification to the single-inclusive charm-jet
  cross section with extrinsic charm only, as a a function of $p_T$
  for three rapidity regions $|y|$.}
\end{figure*}
content at the scale $\mu_F =1.3$ GeV for the BHPS model and the SEA
model of high strength. These two particular models for intrinsic
charm have been employed also for making predictions for the inclusive
production of $D^0$ mesons at $\sqrt{S} = 7$ TeV (to be compared with
LHC data) in \cite{6}. From this work it is clear that the effect of
intrinsic charm as parametrized by the valence-like model BHPS
increases with growing $p_T$ and rapidity. For the sea model SEA the
effect of intrinsic charm occurred uniformly for all $p_T$ and
increased with increasing rapidity $y$ \cite{25}.  We expect a similar
behavior also for charm-jet production. In Fig.~4, we show our results
for the relative enhancement of the $p_T$-distribution in the bins of
$y$ considered previously. The results in Fig.~4 on the left-hand side
are for the $3.5\%$ BHPS model and in Fig.~4 on the right-hand side
for the high-strength sea model. For the BHPS model, $p_T$ must be
large enough to see an increase. For the largest $|y|$ bin: $2.2 \leq
|y| \leq 3.2$, one needs $p_T \geq 80$ GeV to have a rise larger than
two. In the case of the lowest $|y|$ bin: $0.0 \leq |y| \leq 0.5$, the
enhancement is smaller and increases less with growing $p_T$. For the
sea model plotted on the right-hand side of Fig.~4, the enhancement is
nearly independent of $p_T$ and $|y|$ bins and varies between 1.6 and
2.0. Hence, apart from the $|y|$ bin: $0.0 \leq |y| \leq 0.5$, it is
less than for the BHPS model in the area of high $p_T$. For better
visibility the curves for $2.0 \leq |y| \leq 2.2$ and $2.2 \leq |y|
\leq 3.2$ are multiplied by factors 1.25 and 1.5,
respectively. Compared to the inclusive D-meson cross section the
increase seems to be similar if one considers the same $p_T$ and $|y|$
regions. It is clear that for larger $|y|$ it would be bigger. The
question is though, whether these cross sections could be measured
with sufficient accuracy.

Of course, these enhancements in the cross sections would be detectable
only if they are much larger than any variation of the theoretical
cross section due to scale variations. We have not calculated these
scale variations for the charm-jet cross sections, but assume that
they are very similar to the scale variation of the inclusive jet
cross section for the same $p_T$ bin. At the largest $p_T$ bin in
Fig.~4 (174-196 GeV) the scale variation is $13\%(-12\%)$ which is
indeed small compared to the enhancement at this $p_T$ bin, except for
the cross section for the $|y| \leq 0.5$ bin in the case of the
valence-like intrinsic charm (see Fig.4 left).

\section{Summary and Conclusions}

We have calculated the inclusive charm-jet cross section at NLO of QCD
in the zero-mass variable-flavor-number scheme, i.e. with active charm
quarks in the proton, for $pp$ collisions at $\sqrt{S} = 7$ TeV. To
test the presence of intrinsic charm contributions in the proton PDF,
we have employed two PDF sets which are consistent with global PDF
analysis results. We found that they lead to enhancements of the
charm-jet cross section at large transverse momentum and rapidity,
which might possibly be tested with present and future LHC jet
production data.

\section*{Acknowledgements} 

We thank Michael Klasen and Hubert Spiesberger for advice and fruitful
discussions. This work was supported by the German Federal Ministry
for Education and Research BMBF through Grant No.\ 05~H12GUE, by the
German Research Foundation DFG through Grant No. KN 365/5-3, and by
the Helmholtz Association HGF through Gran No. HA 101. IB acknowledges
support by the German Science Foundation DFG through the Collaborative
Research Centre 676 ``Particles, Strings and the Early Universe''.


\begin{thebibliography}{99}

\bibitem{1} S. Frixione, M.L. Mangano, P. Nason et al.,  Adv.\ Ser.\ Direct.\ 
  High Energy Phys.\  {\bf 15} (1998) 609 [hep-ph/9702287].

\bibitem{2} M. Klasen and G. Kramer, Eur.\ Phys.\ J.\ C {\bf 71} (2011) 1774
  [arXiv:1104.0095 [hep-ph]].

\bibitem{3} M. Cacciari and M. Greco,  Nucl.\ Phys.\ B {\bf 421} (1994) 530
  [hep-ph/9311260].

\bibitem{4} J. Binnewies, B.A. Kniehl and G. Kramer, Phys.\ Rev.\ D {\bf 58} 
  (1998) 014014 [hep-ph/9712482] and earlier papers quoted there.

\bibitem{5} B. A. Kniehl, G. Kramer, I. Schienbein and H. Spiesberger,   
  Phys.\ Rev.\ D {\bf 71} (2005) 014018 [hep-ph/0410289] and earlier papers 
  quoted there.

\bibitem{6} B.A. Kniehl, G. Kramer, I. Schienbein and H. Spiesberger,  
  Eur.\ Phys.\ J.\ C {\bf 72} (2012) 2082 [arXiv:1202.0439 [hep-ph]].

\bibitem{7} B. Abelev et al. [ALICE Collaboration],  JHEP {\bf 1201} (2012) 128
  [arXiv:1111.1553 [hep-ex]].

\bibitem{8} The ATLAS Collaboration, ATL-PHYS-PUB-2011-012; ATLAS-CONF-2011-017.

\bibitem{9} R. Aaij et al. [LHCb Collaboration], Nucl.\ Phys.\ B {\bf 871} 
  (2013) 1 [arXiv:1302.2864 [hep-ex]].

\bibitem{10} S. Chatrchyan et al. [CMS Collaboration], Phys.\ Rev.\ Lett.\  
  {\bf 107} (2011) 132001 [arXiv:1106.0208 [hep-ex]].

\bibitem{10a} G. Aad et al. [ATLAS Collaboration], Phys.\ Rev.\ D {\bf 86} 
  (2012) 014022 [arXiv:1112.6297 [hep-ex]]. 

\bibitem{11} M. Klasen and G. Kramer,   Z.\ Phys.\ C {\bf 72} (1996) 107
  [hep-ph/9511405]; ibid. C {\bf 76} (1997) 67 [hep-ph/9611450]; 
  M. Klasen, T. Kleinwort and G. Kramer,  Eur.\ Phys.\ J.\ direct C {\bf 1} 
  (1998) 1 [hep-ph/9712256].

\bibitem{12} M. Klasen,  Rev.\ Mod.\ Phys.\  {\bf 74} (2002) 1221
  [hep-ph/0206169].

\bibitem{13} Z. Nagy,  Phys.\ Rev.\ Lett.\  {\bf 88} (2002) 122003
  [hep-ph/0110315]; Z. Nagy,  Phys.\ Rev.\ D {\bf 68} (2003) 094002
  [hep-ph/0307268].

\bibitem{14} T. Kluge, K. Rabbertz and M. Wobisch in Proceedings of the 14th 
 International Workshop on Deep Inelastic Scattering (DIS 2006), Tsukuba, Japan 
 2006 (World Scientific, Singapore, 2007) p. 483, hep-ph/0609285.

\bibitem{15} H.-L. Lai et al., Phys.\ Rev.\ D {\bf 82} (2010) 074024
  [arXiv:1007.2241 [hep-ph]].

\bibitem{16} S. Chatrchyan et al. [CMS Collaboration], JHEP {\bf 1204} (2012) 
  084 [arXiv:1202.4617 [hep-ex]].

\bibitem{17} P. M. Nadolsky, H.-L. Lai, Q.-H. Cao, J. Huston, J. Pumplin, D. 
 Stump, W.-K. Tung and C. P. Yuan,  Phys.\ Rev.\ D {\bf 78} (2008) 013004
  [arXiv:0802.0007 [hep-ph]].

\bibitem{18} S. Catani et al.,  Nucl.\ Phys.\ B {\bf 406} (1993) 187; 
  S.D. Ellis and D.E. Soper,  Phys.\ Rev.\ D {\bf 48} (1993) 3160
  [hep-ph/9305266].

\bibitem{19} M. Cacciari, G.P. Salam and G. Soyez,  JHEP {\bf 0804} (2008) 063
  [arXiv:0802.1189 [hep-ph]].

\bibitem{20} http://hepdata.cedar.ac.uk/view/ins902309.

\bibitem{22} J. Pumplin, H. L. Lai and W. K. Tung, Phys.\ Rev.\ D {\bf 75} 
  (2007) 054029 [hep-ph/0701220].

\bibitem{23} S. Dulat et al., Phys.\ Rev.\ D {\bf 89} (2014) 073004
  [arXiv:1309.0025 [hep-ph]].

\bibitem{24} LHAPDF (The Les Houches Accord PDF Interface), 
 http://projects.hepforge.org/lhapdf/pdfsets.

\bibitem{25} See also: B. A. Kniehl, G. Kramer, I. Schienbein and 
 H. Spiesberger, Phys.\ Rev.\ D {\bf 79} (2009) 094009
  [arXiv:0901.4130 [hep-ph]]. 

\end{thebibliography}
\end{document}